# Selective Molecular Sieving through Porous Graphene

Steven P. Koenig, Luda Wang, John Pellegrino, and J. Scott Bunch*

*Department of Mechanical Engineering, University of Colorado, Boulder, CO 80309 USA*

*email: jbunch@colorado.edu

**Membranes act as selective barriers and play an important role in processes such as cellular compartmentalization and industrial-scale chemical and gas purification. The ideal membrane should be as thin as possible to maximize flux, mechanically robust to prevent fracture, and have well-defined pore sizes to increase selectivity. Graphene is an excellent starting point for developing size-selective membranes[1–8] because of its atomic thickness[9], high mechanical strength[10], relative inertness, and impermeability to all standard gases[11–14]. However, pores that can exclude larger molecules, but allow smaller molecules to pass through have to be introduced into the material. Here we show UV-induced oxidative etching[15,16] can create pores in micrometre-sized graphene membranes and the resulting membranes used as molecular sieves. A pressurized blister test and mechanical resonance is used to measure the transport of a variety of gases ($H_2$, $CO_2$, Ar, $N_2$, $CH_4$, and $SF_6$) through the pores. The experimentally measured leak rates, separation factors, and Raman spectrum agree well with models based on effusion through a small number of angstrom-sized pores.**

Suspended graphene membranes were fabricated by mechanical exfoliation of graphene over predefined 5 μm diameter wells etched into silicon oxide[17,18]. After exfoliation, the pristine graphene flakes that span the microcavity form suspended membranes that are impermeable to all standard gas molecules[11] and clamped to the silicon oxide substrate by surface forces[18]. Gas species can enter and exit the microcavity through the substrate by slow diffusion. To fill the microcavity with a desired gas species, the sample is put in a chamber pressurized to 200 kPa above ambient pressure with a "charging" gas (Fig. 1a). Prior to this pressurization, the chamber is flushed with the "charging" gas to exclude any other species. The samples are left in the pressure chamber for 4-12 d (depending on the gas species used) to allow for the internal, $p_{int}$, and external pressure, $p_{ext}$, of the microcavity to equilibrate to the "charging" pressure, $p_0$. Upon removing the sample from the pressure chamber the higher pressure inside the microcavity compared with ambient atmospheric pressure causes the membrane to bulge upward (Fig. 1b). This technique allows preparation of a graphene-sealed microcavity with an arbitrary gas composition at a prescribed pressure.

To measure the leak rate of gas species we used both a pressurized blister test and mechanical resonance test[11]. The pressurized blister test was used for leak rates on the order of minutes to hours while the mechanical resonance was used to measure leak rates on the order of seconds to minutes. For the pressurized blister test, an atomic force microscope (AFM) is used to measure the shape of the bulged graphene membrane, which is parameterized by its maximum deflection, $\delta$ (Fig. 1e). The maximum deflection, $\delta$, vs. time, $t$, for a pristine graphene membrane pressurized to 200 kPa, above atmospheric pressure, of $H_2$ gas is shown in Fig. 1f (black). The deflection decreases slowly with time consistent with a leak of $H_2$ gas through the underlying silicon oxide[11,18].

UV-induced oxidative etching was used to introduce pores in the pristine graphene membranes[15,16,19,20] (see supporting online text). The $H_2$ gas pressurized graphene membranes were exposed to UV light ($\lambda_1$ = 185 nm, $\lambda_2$ = 254 nm; Jelight Model 42 UV ozone cleaner) at ambient conditions for several minutes. A number of other etching techniques have been proposed and demonstrated on graphene[19,21–27] but the UV oxidative etching used here is simple and slow enough to allow for the creation of these subnanometer-sized selective pores as demonstrated later in this paper. Other etching techniques, including oxygen plasma etching, were tried but UV oxidative etching proved to be the only successful method for controllably introducing subnanometer pores. After the oxidative etch, $\delta$ is again measured versus $t$ (Fig. 1e and 1f, red) (see supplementary info). The maximum deflection decreases rapidly (several minutes as opposed to hours for the unetched case) and eventually leads to a downward deflection of the membrane (Fig. 1c-1f). Figure 1e shows a series of cross sections through the centre of the membrane taken by AFM as time elapses from 0 to 8 min and Fig 1g shows a three dimensional rendering of the AFM image for $t$=0 in Fig. 1e. Here 0 min is defined to be the time at which the first AFM image was captured after removing the sample from the pressure chamber. The change in deflection, as depicted in Fig. 1c & d, results from increasing the $H_2$ leak rate, through etching, while preventing significant changes in the $N_2$ leak rate into the microcavity from the ambient atmosphere.

The molecular selectivity of the fabricated porous graphene membrane is demonstrated by measuring the time rate of change of $\delta$, $-d\delta/dt$, for the same membrane pressurized with a number of different gases. Figure 2a shows $\delta$ vs. $t$ for $H_2$, $CO_2$, Ar, and $CH_4$ before and after etching and $N_2$ after etching. We did not measure the $N_2$ leak rate for this particular device before etching, but measurements for 12 other ones located on the same flake are shown in Fig. 4 and labelled "Pristine Avg" for comparison with the after-etch leak rate. At short times, $-d\delta/dt$ is approximately linear (Fig. 2a). This rate, $-d\delta/dt$, versus kinetic diameter[28] is plotted for all the gases measured for the same membrane/microcavity in Fig. 1 before and after etching (Fig. 2b). After etching, there is an increase in $-d\delta/dt$ of two orders of magnitude for the leak rate of $H_2$ and $CO_2$, while Ar and $CH_4$ remain relatively unchanged. This suggests that the etched pores change the transport mechanism for $H_2$ and $CO_2$, while leaving the transport of Ar and $CH_4$ nearly unchanged. Since the kinetic diameter cut off in this bi-layer graphene membrane is nominally that of Ar, 3.4 Å[28], this membrane will heretofore be referred to as "Bi- 3.4 Å".

The leak rate of various gases across the porous graphene membranes can also be measured by using a mechanical resonance test. This is accomplished by measuring changes in the mechanical resonant frequency, $f$, of the membrane vs. $t$ using an optical drive and detection system previously used to measure mechanical resonance in suspended graphene resonators[11,29]. A pressure difference applied across the membrane leads to a pressure-induced tensioning of the membrane, which increases $f$ of the stretched membrane. If the gas molecules introduced external to an initially evacuated microcavity can leak through the membrane, the gas will pass through and reduce the tension in the membrane, thus decreasing $f$. If the gas molecules cannot leak through the membrane, $f$ stays constant. An example of this is shown in Fig. 3 where an etched porous graphene membrane was put in a vacuum of 0.1 torr for a several days to ensure the microcavity has equilibrated to the pressure of the vacuum chamber. Next, a pure gas species is introduced into the vacuum chamber at a given pressure (~100 torr for the case in Fig. 3 and ~80 torr for the inlay of Fig. 3) and the resonant frequency is measured. The resonant frequency decreases with time, and from the rate of decrease, we determine the leak rate through porous graphene membrane. We could not observe the frequency return back to its original value due to significant gas damping when $\Delta p$ ~ 0 (see supporting online text). As can be seen from Fig. 3, the leak rate of $H_2$, $CO_2$, $N_2$, and $CH_4$ is several seconds while $SF_6$ shows no significant change in resonant frequency for the several minutes measured. This membrane will be referred to as “Bi- 4.9 Å” since it is a bilayer membrane with a nominal sieving kinetic diameter of $SF_6$, 4.9 Å[28].

We derived the following expression for the molecular flux out of the pressurized "blister" microcavity, *dn/dt*, using the ideal gas law and Hencky’s solution for a clamped circular membrane[30] (see supplementary info for derivation):

$$\frac{dn}{dt} = \frac{3\mathrm{K}\ \nu\ \ (Ew\delta^2)/a^4 \cdot V(\delta) + P(C(\nu)\pi a^2)}{\mathrm{R}T} \cdot \frac{d\delta}{dt} \qquad (1)$$

where *a* is the radius of the membrane, *E* is the Young’s modulus, *w* is the thickness of the membrane, R is the molar gas constant, *T* is temperature, *V(δ)* is the total volume of the microcavity in the bulged state, and *C(v)* and *K(v)* are geometric coefficients which depend on the Poisson’s ratio, *v*, of the membrane. For the case of graphene, the Young’s modulus and Poisson’s ratio are $E$ = 1 TPa and $v$ = 0.16, respectively, and the thickness per layer is 0.34 nm[10,11,18,31]. Using $v$ = 0.16 gives coefficients of $K(v=0.16)$ = 3.09 and $C(v=0.16)$ = 0.524[17]. Figure 4 shows the normalized *dn/dt* (normalized to the partial pressure difference across the membrane) for the “Bi- 3.4 Å” membrane before UV etching (black squares) and after UV etching (red squares). Also included is the average normalized *dn/dt* for 24 different unetched (12 for the case of $N_2$) membranes on the same graphene flake shown in the Fig 1f inlay that contains “Bi- 3.4 Å” (black circles). Similarly, *dn/dt*, can be calculated from the linear approximation of the rate of frequency decay, *df/dt* (see supplementary info for details). The leak rate versus molecular size for the “Bi- 4.9 Å” membrane is shown in Fig. 4a (red diamonds).

The changes in leak rates associated with UV etching are consistent with the introduction of a pore(s) which allow size selective permeation of gas molecules. For the “Bi- 3.4 Å” membrane in Fig. 2, the selectivity between $CO_2$ and Ar suggests that the pore(s) size(s) introduced into the graphene membrane are comparable to the kinetic diameter of Ar (3.4 Å)[28] and that the porous graphene is sieving molecules above and below this size. Similarly for the “Bi- 4.9 Å” membrane in Fig. 3, there are likely pore(s) larger in size than that of the “Bi- 3.4 Å” membrane, since effective molecular sieving is seen for molecules smaller than $SF_6$ (4.9 Å compared to 3.8 Å for $CH_4$)[28]. Due to the fact that there is likely only a small density of pores in the 5 μm diameter membranes, imaging of the pore is not possible (see supporting online text). However, the small density of pores is supported by Raman spectroscopy on the etched membranes (see supporting online text).

The gas leak rates measured can be compared to results of computational modelling by Jiang *et al.* and Blankenburg *et al.*[1,5]. Following the work of Jiang *et al.*, we estimate a $H_2$ leak rate on the order of ~$10^{-20}$ mol $s^{-1}$ $Pa^{-1}$ for a H-passivated pore in graphene consisting of 2 missing benzene rings at room temperature (see supporting online text)[1]. For the work of Blankenburg *et al.*, the $H_2$ leak rate was calculated to be on the order of ~$10^{-23}$ mol $s^{-1}$ $Pa^{-1}$ through a smaller H-terminated pore consisting of a single missing benzene ring[5].

Our measured $H_2$ leak rate on “Bi- 3.4 Å” was ~4.5 x $10^{-23}$ mol $s^{-1}$ $Pa^{-1}$. This value is several orders of magnitude lower than Jiang *et al.*, suggesting our pores have an overall higher energy barrier for $H_2$ (and other species) than in their calculations. The similarity between our $H_2$ leak rate with that modelled by Blankenburg *et al.* suggests a similar $H_2$ energy barrier in our pore. Nonetheless, we do not match their calculated $H_2/CO_2$ selectivity (2 versus ~$10^{17}$). This suggests that having a bilayer graphene membrane with different chemical pore termination from the oxidative etching can be quite important.

We can also compare the $H_2$ and $CO_2$ measured leak rates between the “Bi- 3.4 Å” and “Bi- 4.9 Å” membranes (Fig. 4). The one with the smaller pore size, “Bi- 3.4 Å”, (red squares) had a $H_2$ and $CO_2$ leak rates (in units of $10^{-23}$ mol $s^{-1}$ $Pa^{-1}$) of 4.5 and 2.7, respectively, compared to $H_2$ and $CO_2$ leak rates (same units) of 75 and 25, respectively, for the larger pore membrane (red diamonds). The closeness between the magnitude of these 2 values, and the magnitudes calculated in the cited modelling, suggests that in both cases a low density of size-selective pores are participating in the transport across the graphene membrane and the faster leak rate for the “Bi- 4.9 Å” membrane is consistent with larger pores (and/or lower diffusional energy barriers) than the “Bi- 3.4 Å”. This is also consistent with the rapid effusion of gas expected from the ~$\mu m^3$ confined volume of gas in the porous graphene sealed microchamber[11].

Both graphene membranes presented here were bilayer graphene membranes due to the more controlled etching and stability of the pores fabricated on bilayer versus monolayer graphene membranes. This is consistent with previous results showing slower etching for bilayer graphene compared with single layer graphene[19]. However, similar results were observed on monolayer graphene membranes (see supporting online text).

In conclusion, we have demonstrated selective molecular sieving using porous, μm-sized, atomically-thin graphene membranes. Pores were introduced in graphene by UV-induced oxidative etching and the molecular transport through them was measured using both a pressurized blister test and mechanical resonance. Our results are consistent with theoretical models in the literature based on effusion through angstrom-sized pores[1,5]. The results presented here are an experimental realization of graphene gas separation membranes by molecular sieving and represent an important step towards the realization of macroscopic, size-selective porous graphene membranes. The approach used here can also be used to probe the fundamental limits of gas transport by effusion through angstrom-sized pores with atomic-sized channel lengths.

**Methods**

Suspended graphene membranes are fabricated by a combination of standard photolithography and mechanical exfoliation of graphene. First, an array of circles with diameters of 5 μm and 7 μm are defined by photolithography on an oxidized silicon wafer with a silicon oxide thickness of 285 nm. Reactive ion etching is then used to etch the circles into cylindrical cavities with a depth of 250-500 nm leaving a series of wells on the wafer. Mechanical exfoliation of Kish graphite using Scotch® tape is then used to deposit suspended graphene sheets over the wells.

The volume of the bulged graphene is on the order of the initial volume of the microcavity[17], and we deduce the initial $\Delta p = p_{int} - p_{ext}$ across the membrane, using the ideal gas law and isothermal expansion of the trapped gas with a constant number of molecules, N. Doing so leads to $p_o V_o = p_{int}(V_o+V_b)$, where $V_o$ is the initial volume of the well and $V_b = C(\upsilon)\pi a^2\delta$ is the volume of the pressurized blister after the device is brought to atmospheric pressure and bulges upward. The constant, $C(\upsilon = 0.16) = 0.524$ is determined from Hencky's solution. AFM scans are then continuously taken in order to deduce the leak rate of molecules out of the membrane, $dn/dt$.

For the resonance measurements, samples are placed in a vacuum chamber at 0.1 torr for several days to ensure the microcavity reaches equilibrium with the vacuum chamber. A given pressure (ranging from 80-100 torr) of gas is then introduced into the vacuum chamber and the frequency is measured over time. After the introduction of a gas, the chamber is evacuated until the frequency returns to its original value when no pressure difference was present (or the signal was no longer detectable due to gas damping) and the next gas is then measured. An intensity modulated blue laser (405nm) was used to drive the graphene membranes and a red laser (633nm) was used to detect the motion of the graphene.

## References

1. Jiang, D., Cooper, V.R. & Dai, S. Porous graphene as the ultimate membrane for gas separation. *Nano Lett.* **9**, 4019–4024 (2009).

2. Du, H. *et al.* Separation of Hydrogen and Nitrogen Gases with Porous Graphene Membrane. *J. Phys. Chem. C* **115**, 23261–23266 (2011).

3. Schrier, J. Helium Separation Using Porous Graphene Membranes. *J. Phys. Chem. Lett.* **1**, 2284–2287 (2010).

4. Hauser, A.W. & Schwerdtfeger, P. Nanoporous Graphene Membranes for Efficient 3 He/ 4 He Separation. *J. Phys. Chem. Lett.* **3**, 209–213 (2012).

5. Blankenburg, S. *et al.* Porous graphene as an atmospheric nanofilter. *Small* **6**, 2266–2271 (2010).

6. Suk, M.E. & Aluru, N.R. Water Transport through Ultrathin Graphene. *J. Phys. Chem. Lett.* **1**, 1590–1594 (2010).

7. Schrier, J. & McClain, J. Thermally-driven isotope separation across nanoporous graphene. *Chem. Phys. Lett.* **521**, 118–124 (2012).

8. Li, Y., Zhou, Z., Shen, P. & Chen, Z. Two-dimensional polyphenylene: experimentally available porous graphene as a hydrogen purification membrane. *Chem. Commun.* **46**, 3672–3674 (2010).

9. Meyer, J.C. *et al.* The structure of suspended graphene sheets. *Nature* **446**, 60–63 (2007).

10. Lee, C., Wei, X., Kysar, J.W. & Hone, J. Measurement of the Elastic Properties and Intrinsic Strength of Monolayer Graphene. *Science* **321**, 385–388 (2008).

11. Bunch, J.S. *et al.* Impermeable atomic membranes from graphene sheets. *Nano Lett.* **8**, 2458–2462 (2008).

12. Nair, R.R., Wu, H.A., Jayaram, P.N., Grigorieva, I.V. & Geim, A.K. Unimpeded Permeation of Water Through Helium-Leak-Tight Graphene-Based Membranes. *Science* **335**, 442–444 (2012).

13. Leenaerts, O., Partoens, B. & Peeters, F.M. Graphene: A perfect nanoballoon. *Appl. Phy. Lett.* **93**, 193107 (2008).

14. Chen, S. *et al.* Oxidation Resistance of Graphene-Coated Cu and Cu/Ni Alloy. *ACS Nano* **5**, 1321–1327 (2011).

15. Ozeki, S., Ito, T., Uozumi, K. & Nishio, I. Scanning Tunneling Microscopy of UV-Induced Gasification Reaction on Highly Oriented Pyrolytic Graphite. *Jpn. J. Appl. Phys.* **35**, 3772–3774 (1996).

16. Huh, S. *et al.* UV/Ozone-Oxidized Large-Scale Graphene Platform with Large Chemical Enhancement in Surface-Enhanced Raman Scattering. *ACS Nano* **5**, 9799–9806 (2011).

17. Novoselov, K.S. *et al.* Two-dimensional atomic crystals. *Proc. Natl Acad. Sci. USA* **102**, 10451–10453 (2005).

18. Koenig, S.P., Boddeti, N.G., Dunn, M.L. & Bunch, J.S. Ultrastrong adhesion of graphene membranes. *Nature Nanotech.* **6**, 543–546 (2011).

19. Liu, L. *et al.* Graphene oxidation: thickness-dependent etching and strong chemical doping. *Nano Lett.* **8**, 1965–1970 (2008).

20. Chang, H. & Bard, A.J. Scanning tunneling microscopy studies of carbon-oxygen reactions on highly oriented pyrolytic graphite. *J. Am. Chem. Soc.* **113**, 5588–5596 (1991).

21. Bieri, M. *et al.* Porous graphenes: two-dimensional polymer synthesis with atomic precision. *Chem. Commun.* 6919–6921 (2009).

22. Girit, C.O. *et al.* Graphene at the edge: stability and dynamics. *Science* **323**, 1705–1708 (2009).

23. Schrier, J. Fluorinated and Nanoporous Graphene Materials As Sorbents for Gas Separations. *ACS Appl. Mater. Interfaces* **3**, 4451–4458 (2011).

24. Bai, J., Zhong, X., Jiang, S., Huang, Y. & Duan, X. Graphene nanomesh. *Nature Nanotech.* **5**, 190–194 (2010).

25. Sint, K., Wang, B. & Král, P. Selective ion passage through functionalized graphene nanopores. *J. Am. Chem. Soc.* **130**, 16448–16449 (2008).

26. Fan, Z. *et al.* Easy synthesis of porous graphene nanosheets and their use in supercapacitors. *Carbon* **50**, 1699–1703 (2012).

27. Fox, D. *et al.* Nitrogen assisted etching of graphene layers in a scanning electron microscope. *Appl. Phys. Lett.* **98**, 243117 (2011).

28. Breck, D.W. *Zeolites Molecular Sieves: Structure, Chemistry, and Use*. 593–724 (Wiley: New York, NY, 1973).

29. Bunch, J.S. *et al.* Electromechanical Resonators from Graphene Sheets. *Science* **315**, 490–493 (2007).

30. Hencky, H. Uber den spannungzustand in kreisrunden platten mit verschwindender biegungssteiflgeit. *Z. fur Mathematik und Physik* **63**, 311–317 (1915).

31. Blakslee, O.L., Proctor, D.G., Seldin, E.J., Spence, G.B. & Weng, T. Elastic Constants of Compression-Annealed Pyrolytic Graphite. *J.Appl. Phys.* **41**, 3373–3382 (1970).

## Acknowledgements

We thank Darren McSweeney and Michael Tanksalvala for help with the resonance measurements and Rishi Raj for use of the Raman microscope. This work was supported by NSF Grants #0900832(CMMI: Graphene Nanomechanics: The Role of van der Waals Forces), #1054406 (CMMI: CAREER: Atomic Scale Defect Engineering in Graphene Membranes), the DARPA Center on Nanoscale Science and Technology for Integrated Micro/Nano-Electromechanical Transducers (iMINT), the National Science Foundation (NSF) Industry/University Cooperative Research Center for Membrane Science, Engineering and Technology (MAST), and by the NNIN and the National Science Foundation under Grant No. ECS-0335765.

## Author Contributions

## Additional Information

The authors declare no competing financial interest. Supplementary information accompanies this paper at www.nature.com/naturenanotechnology. Reprints and permission information is available online at http://npg.nature.com/reprintsandpermissions/. Correspondence and requests for materials should be addressed to J.S.B.

## Figure Captions

### Figure 1: Measuring Leak Rates in Porous Graphene Membranes

**(a)** Schematic of a microscopic graphene membrane on a silicon oxide substrate. We start with pristine graphene fabricated by exfoliation and fill the microchamber with 200 kPa of $H_2$ (represented as red circles here) in a pressure chamber. Equilibrium is reached ($p_{int} = p_{ext}$) by diffusion through the silicon oxide.

**(b)** After removing the graphene membrane from the pressure chamber the membrane is bulged upward. We calculate $p_{int}$ using the ideal gas law and assuming isothermal expansion. The hydrogen molecules slowly leak out of the microchamber through the silicon oxide substrate.

**(c)** Upon etching of the graphene membrane pore(s) bigger than that of $H_2$ are introduced allowing the $H_2$ to leak rapidly out of the microchamber through the graphene membrane. If the pore(s) are smaller than that of air molecules (mostly $N_2$ and $O_2$, denoted as green circles), air will be blocked from entering the microchamber causing the deflection of the graphene membrane to continue to decrease until all of the $H_2$ molecules exited the microchamber.

**(d)** After all the $H_2$ molecules have leaked out of the microchamber the membrane will be bulged downward.

**(e)** Deflection versus position, 0 min (black) through 8 min (dashed blue) after etching, corresponding to some of the red points in (**f**).

**(f)** Maximum deflection, $\delta$, vs. $t$ for one membrane that separates $H_2$ from air as measured by AFM. The black points represent the leak rate of $H_2$ before etching and the red points show the leak rate of $H_2$ after introducing selective pores in the graphene. Inlay: Optical image of the bilayer graphene flake used in this study covering many cavities in the silicon oxide substrate.

**(g)** Three dimensional rendering of an AFM image corresponding to the line cut at $t$ = 0 in (**e**).

**Figure 2: Comparing Leak Rates between Pristine and Porous Graphene Membranes**

**(a)** Maximum deflection, $\delta$, versus $t$ before (black) and after etching (red).

**(b)** Average $-d\delta/dt$ versus molecular size found from the slopes of membrane deflection versus $t$ in (**a**) for before (black) and after (red) introducing pores in the same graphene membrane. The connecting lines show the measurements before (black) and after (red) etching.

**Figure 3: Measuring Leak Rates in a Porous Graphene Membrane Using Mechanical Resonance**

Frequency, *f*, versus *t* for $H_2$ (black), $CO_2$ (red), $N_2$ (green), $CH_4$ (blue), and $SF_6$ (cyan). With a pressure of 100 torr (~13.3 kPa) introduced into the vacuum chamber. Inlay is data from the same device with an 80 torr (~10.7 kPa) pressure introduced.

**Figure 4: Compilation of Measured Leak Rates**

Leak rate out of the microcavity for: "Bi- 3.4 Å" membrane before etching (black squares) and after etching (red squares), "Bi- 4.9 Å" membrane after etching (red diamonds), and the average before etching of 24 membranes (12 for $N_2$) on the same graphene flake as "Bi- 3.4 Å" membrane (black circles with dot). (Note: the latter are hidden by black squares for several gases.)

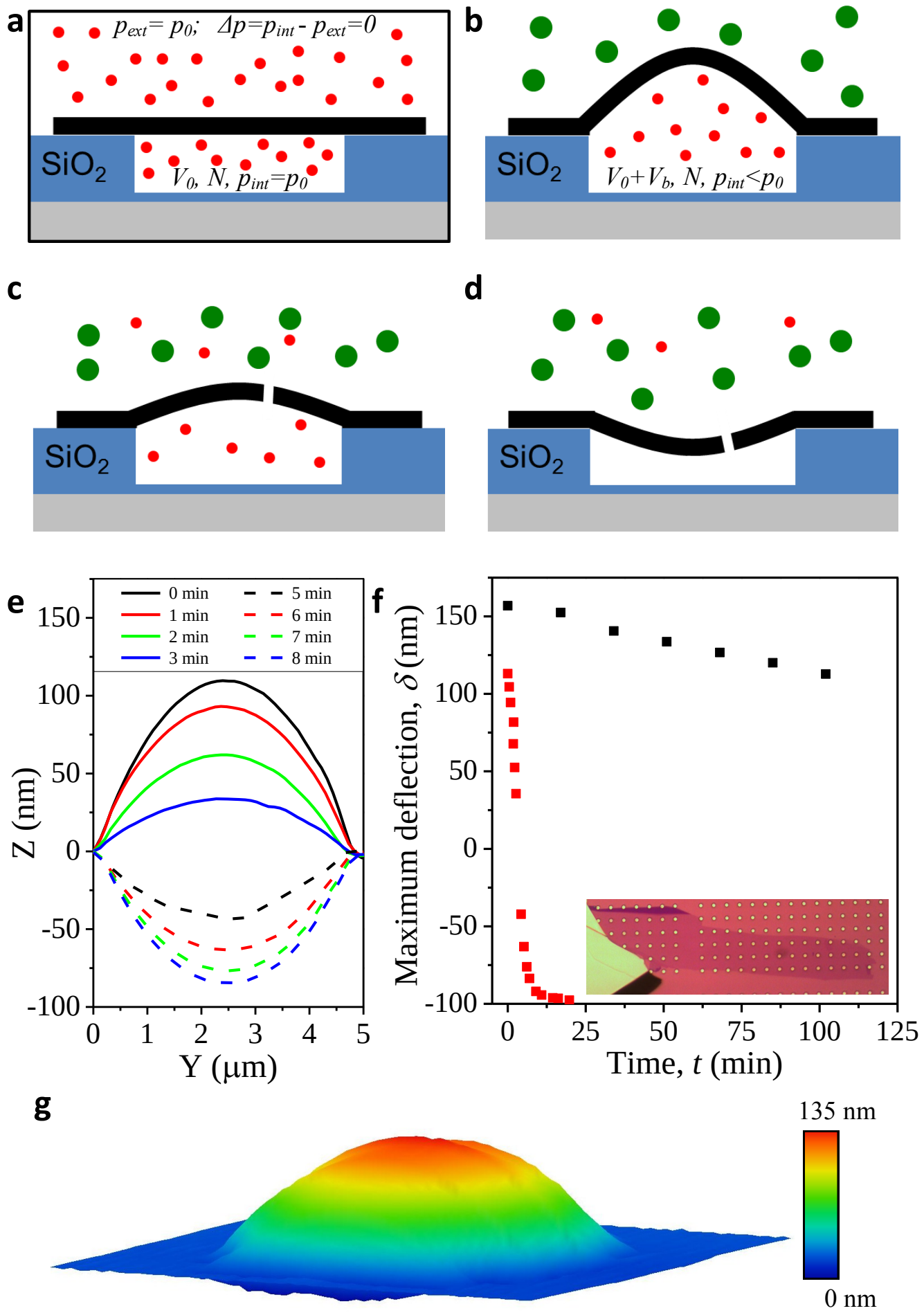

**Figure 1**

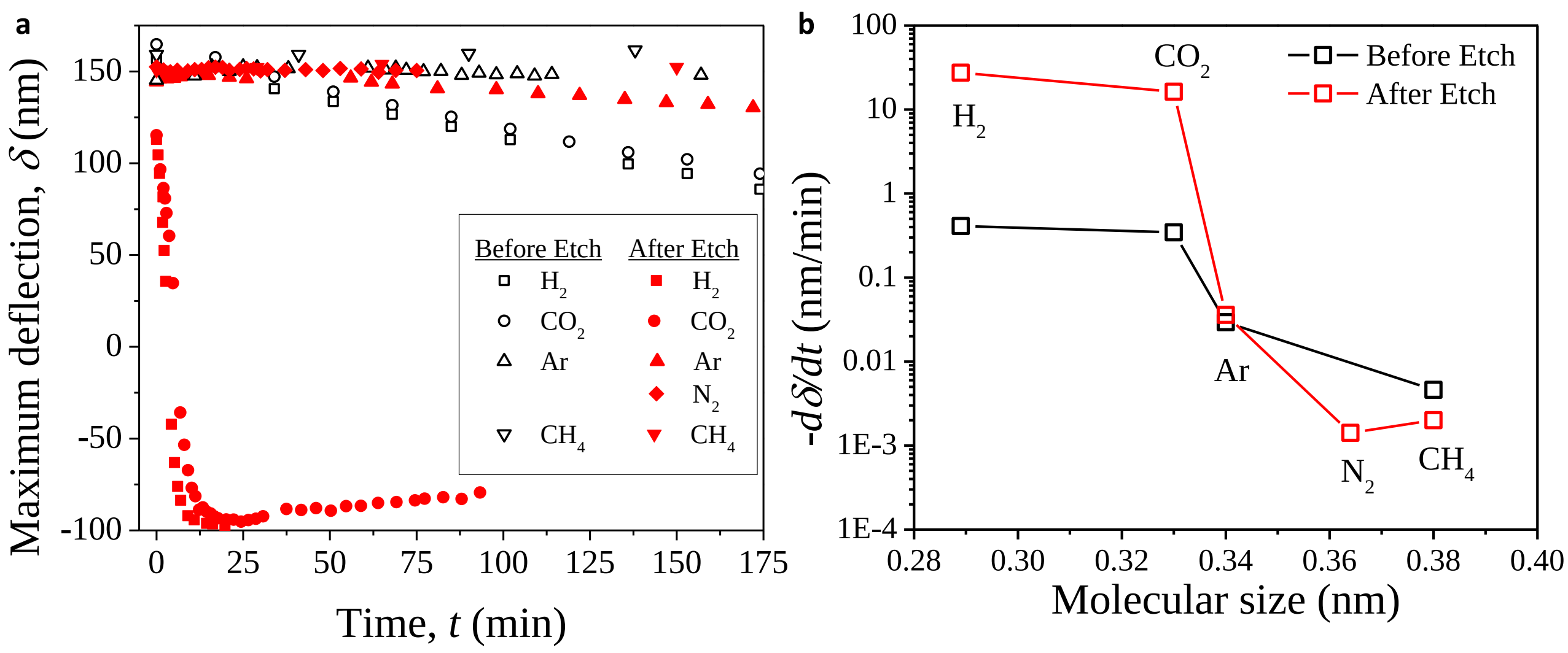

**Figure 2**

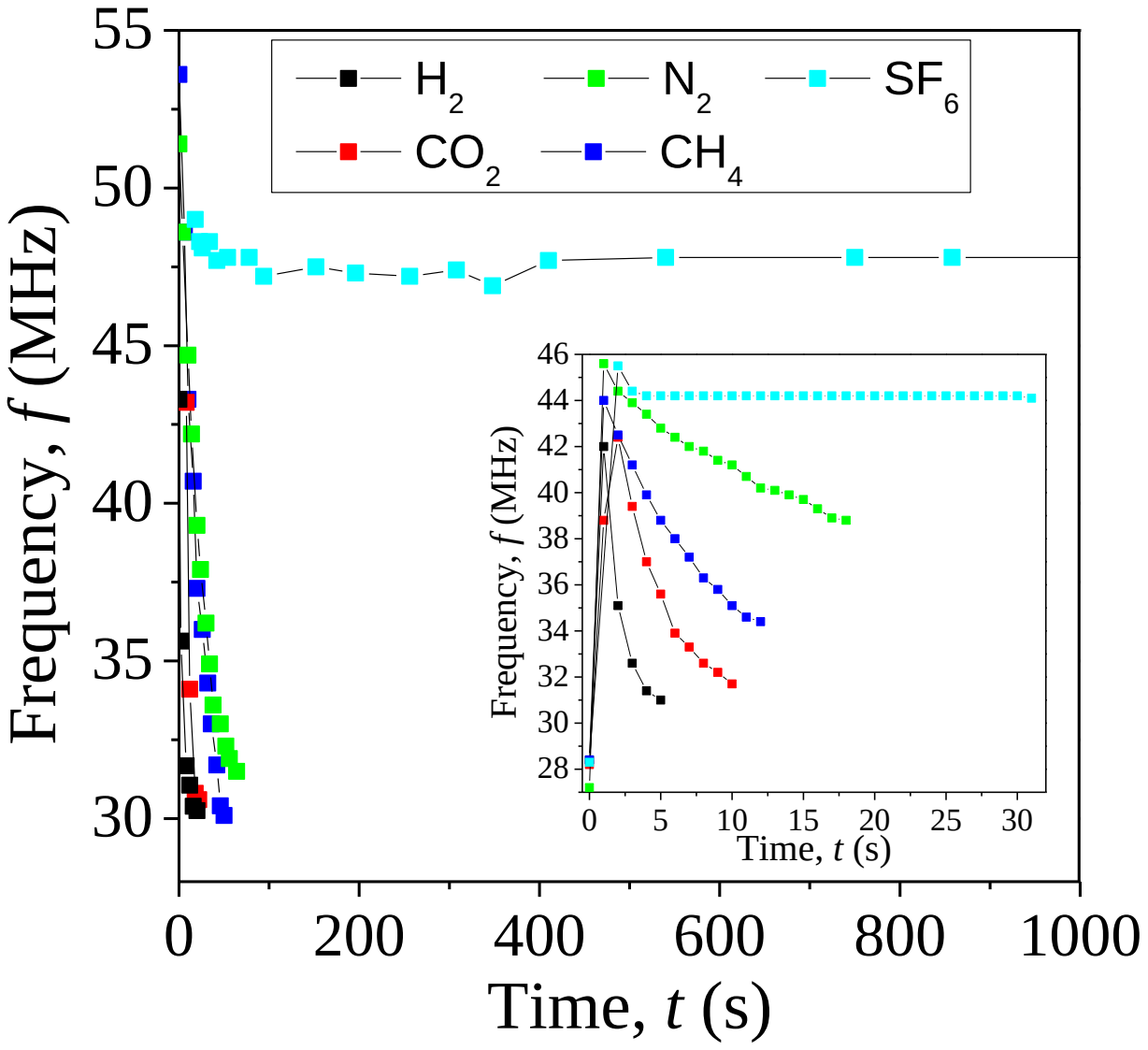

**Figure 3**

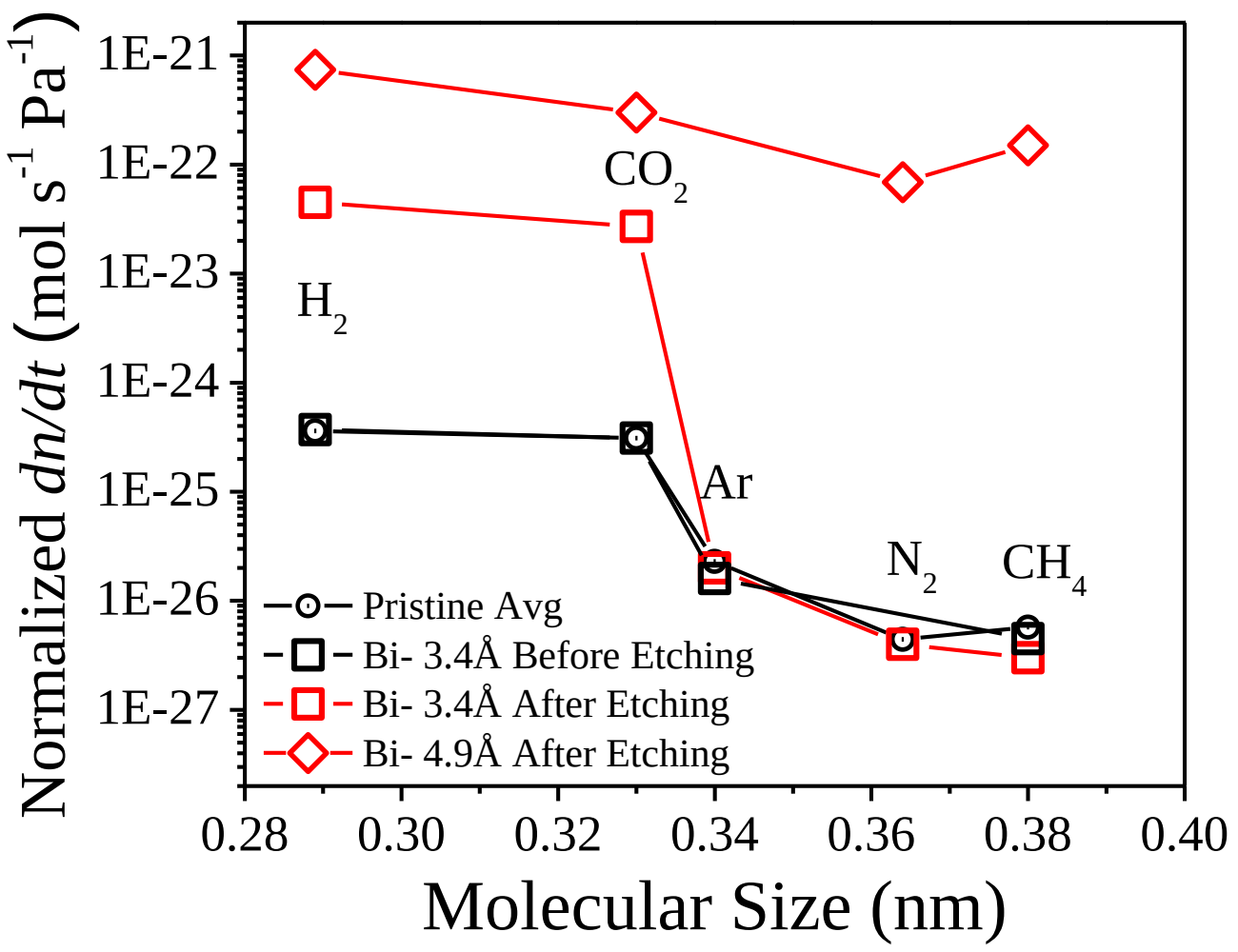

**Figure 4**

# Selective Molecular Sieving Through Porous Graphene

Steven P. Koenig, Luda Wang, John Pellegrino, and J. Scott Bunch*

*email: jbunch@colorado.edu

**Supplementary Information:**

## Raman Spectrum of Graphene Flakes

Raman spectroscopy was used to support our conclusion that a small number of pores exist in the graphene flakes. The D-peak (1360 $cm^{-1}$ wavenumber) is associated with defects in the graphene lattice. Figure S1 shows the Raman spectrum of the graphene flakes used in this study. Figure S1a shows the Raman spectrum of membrane "Bi- 3.4 Å" presented in the main text. This spectrum was taken "before-etching" but is identical to the "after-etching" Raman spectrum. Both spectrums show no D-peak. Figure S1b shows the ratio of the graphene G-peak to the silicon peak areas for the flake presented in this study (the red closed square) and a nearby flake which contained mono- and bi-layer portions (the black open circle for mono-layer and red open square for bi-layer). This follows the work by Koh et al to confirm we had bilayer graphene[1]. Figure S1c shows the Raman spectrum for membrane "Bi- 4.9 Å" presented in the main text "before-etching" showing the characteristic 2D peak shape of bilayer graphene. Figure S1d shows the Raman spectrum of the two monolayer membranes presented in the supplementary information, "Mono-3.4 Å", from Figure S5, (upper red curve) and "Mono-5 Å", presented in Figure S6 (lower black curve). Both spectrums presented in Figure S1d were taken "after-etching". "Mono- 3.4 Å" showed similar $H_2$ leaking behavior as membrane "Bi- 3.4 Å" presented in the main text. No spatial variation was seen in the Raman spectrum of these flakes after-etching. There has been no D-peak observed in etched monolayer samples that showed gas selectivity and we have observed a D-peak in bilayer samples showing selectivity but future work will be needed to correlate the D-peak with pore density since the top layer of bilayer graphene will likely etch before the bottom layer does to open up pores.

## Etching Pores in Graphene Membranes

In order to etch the graphene membranes, we first pressurized them with pure $H_2$ up to 200 kPa (gauge pressure) above ambient pressure. After the microcavity reached equilibrium we removed it from the pressure chamber and measured the deflection using atomic force microscopy (AFM). We then did a series of short UV etches (30 s) followed by AFM scans between each etching step to see if the leak rate increased significantly. When pore(s) were created that were selective to allow the $H_2$ to pass through, but not allow the molecules in the air to pass, the deflection would rapidly decrease and become negative, consistent with a vacuum inside the microcavity. For the case of the “Bi- 3.4 Å” membrane in the main text, this etching took 75 min (150, 30 s etching steps). Each etch step took about 5 min to complete. Once the sample was out of the pressure chamber for over an hour during the etching process, and the deflection had decreased 20 nm, we then returned the sample to the pressure chamber overnight to allow the pressure inside the microchamber to once again reach 200 kPa. The etching process was then continued the next day. For membrane “Bi- 4.9 Å” in the main text, the total etching time was 15 min using 1 min etching steps. From the etching experiments it was noted that longer etch steps required significantly less total etching time.

Since we conclude that there are only a small number of sub-nanometer pores in the 5 μm membranes, direct imaging of these pores is not possible. For classical effusion of gas out of the microcavity, the number of molecules in the microcavity is given by:

$$n = n_0 e^{-\frac{A}{V}\sqrt{\frac{k_b T}{2\pi m}}t} \qquad (S1)$$

where $n_0$ is the initial number of molecules, $A$ is the area of the hole, $V$ is the volume of the container, $k_b$ is Boltzman's constant, $T$ is temperature, $t$ is time, and $m$ is the molecular mass of the gas undergoing effusion[2,3]. For a 3 Å diameter circular pore, and 100 kPa $H_2$ pressure, the leak rate is ~$10^{-20}$ mol·$s^{-1}$·$Pa^{-1}$ which should be fast enough to experimentally measure by our technique and on the order of the leak rates presented here.

In order to visualize pores created by the UV induced oxidative etching reported in the main text, one membrane was over-etched to create much larger pores so we could image the pore formation and distribution with AFM. Figure S2 shows a monolayer membrane that was over-etched (22 min total with 1 min etching steps) in order to visualize the pore growth. Fig S2a shows the 500 nm x 500 nm AFM scan over the suspended region of the over-etched graphene membrane. This membrane was not selective to any of the gas species tested and the leak rates were too fast to measure. The results of the pore size distribution seen in Fig. S2b and Fig. S2c are comparable to previous oxidative etching of graphene and graphite (see references 14,15, and 19 from them main text).

### Calculating the Pressure Normalized Leak Rate from Deflection versus *t* Data

The deformation of the membrane can be described using Hencky's (1915)

solution for a pressurized clamped circular elastic membrane with a pressure difference of $\Delta p$ across it:

$$\Delta p = K(\nu)(Ew\delta^3)/a^4 \qquad (S2)$$

where $E$ is the Young's modulus, $\nu$ is the Poisson's ratio, $w$ is the membrane thickness, and $K(\nu)$ is a coefficient that depends only on $\nu$ [4]. For the case of graphene, we take $E$=1TPa and $\nu$=0.16, therefore $K(\nu=0.16)=3.09$[5]. In order to derive $dn/dt$, the leak rate of the microcavity, we start with the ideal gas law:

$$PV(\delta) = nRT \qquad (S3)$$

where $P$ is the absolute pressure inside the microcavity, $V(\delta)$ is the volume of the microcavity when the membrane is bulged with deflection $\delta$, $V(\delta)=V_o+V_b(\delta)$, $V_b(\delta)=C(\nu)\pi a^2\delta$, for graphene $C(\nu = 0.16) = 0.52$, $n$ is the number of moles of gas molecules contained in the microcavity, $R$ is the gas constant, and $T$ is temperature[5]. Substituting $(\Delta p+p_{atm})$ for $P$ and dividing both sides by $V(\delta)$, and inserting Henckey's solution for $\Delta p$ we get:

$$\left(K(\nu)(Ew\delta^3)/a^4 + p_{atm}\right) = \frac{nRT}{V_o + V_b(\delta)} \qquad (S4)$$

Now we can take the time derivative of both sides and solve for $dn/dt$ to get the flux of gas molecules out of the membrane:

$$\frac{dn}{dt} = \frac{[3K(\nu)(Ew\delta^2)/a^4 \cdot V(\delta) + P(C(\nu)\pi a^2)]}{RT} \cdot \frac{d\delta}{dt} \qquad (S5)$$

To get the $dn/dt$ (mol/s), we use the measured $d\delta/dt$, the rate of the bulge decay from the linear fit of the membrane deflection versus time data. We then normalize the leak rate by dividing the calculated $dn/dt$ by the pressure driving force for each of the gases measured to get the leak rate out of the microcavity.

### Calculating the Pressure Normalized Leak Rate from Frequency versus *t* Data

A schematic of the resonance measurement is presented in Fig S3. Figure S3a shows a membrane that is exposed to a gas smaller than the pore(s) in the graphene thus able to pass through after the membrane has been initially placed in vacuum. Over time, the molecules will leak into the microcavity causing the deflection, and thus the tension, to decrease which leads to a decreasing resonant frequency. Figure S3b shows the membrane in a gas species that is larger than the pore(s) in the graphene. Since the gas is larger than the pore(s) it is blocked and the resonant frequency does not change over time. For the case of the gas being able to pass through the graphene membrane, once the pressure begins to equilibrate on both sides, the signal is lost due to significant gas damping, and it is not possible to accurately experimentally resolve the resonant frequency. This can be seen in the $CH_4$ data presented Fig S4. This data is resonant frequency curves from the $CH_4$ leak rate found in Fig 3 inlay of the main text with 80 torr initially introduced across the membrane. The black curve is the original frequency, $t = 0$ s. The red curve is the frequency right after the pressure is introduced to the membrane, $t = 1$ s. From the red curve you can see there is already a significant gas damping which is evident because of the lower quality factor (i.e. broader peak). The green, blue, cyan, and magenta curves correspond to $t = 3$ s, $t = 5$ s, $t = 7$ s, and $t = 11$ s, respectively. At $t = 13$ s (orange curve) the damping is too large to discern a peak and we cannot determine what the resonant frequency is at or after this time.

The frequency of a circular membrane under tension caused by a pressure difference $\Delta p$ can be described using the following 3 equations:

$$f = \frac{2.404}{2\pi}\sqrt{\frac{S}{\rho_A a^2}} \qquad (S6)$$

$$S = \frac{\Delta p a^2}{4\delta} + S_0 \qquad (S7)$$

$$\Delta p = K(\nu)(Ew\delta^3)/a^4 + (4S_0\delta)/a^2 \qquad (S8)$$

where $f$ is the resonant frequency of the membrane, $a$ is the radius of the membrane, $S$ is the tension in the membrane due to the applied pressure and $S_0$ is the initial tension in the membrane, and $\rho_A$ is the mass density[6]. $K(\nu)$ and $E$ are elastic constants from Hencky's solution and $w$ is the thickness of the membrane and $\delta$ is the deflection in the membrane[4]. We do not take $S_0$ to be zero in this case since the pressure difference and thus the deflection of the membrane are small compared with the case of the blister test. In order to derive the *dn/dt*, the leak rate of the microcavity we first need to solve for $S$ by combining (S7) and (S8) to get:

$$S^3 - 4S^2 + 5S{S_0}^2 - 2{S_0}^3 = K(\nu)(Ewa^2\Delta p^2)/64 \qquad (S9)$$

Since $S$ is larger than $S_0$ we can neglect the cubic order term of $S_0$. Now we can insert the expression for $S$ into equation (S6) and solve for $\Delta p$, and then insert this expression for $\Delta p$ into the ideal gas law in a similar fashion as the bulge test equation. Since the deflection of the membrane is small in this case we take $V$ to be constant. After doing this and taking time derivative and solving for *dn*/*dt* we arrive at the expression:

$$\frac{dn}{dt} = \frac{V}{RT}\left[\left(\frac{(c_1\rho_A - 2c_2{\rho_A}^2a^2f^2){S_0}^2 + c_3{\rho_A}^3a^4f^4}{4K(\nu)Ew}\right)^{1/2} + \frac{c_3{\rho_A}^3a^4f^4 - c_2S_0{\rho_A}^2a^2f^2}{\left(K(\nu)Ew((c_1\rho_A - 2c_2{\rho_A}^2a^2f^2){S_0}^2 + c_3{\rho_A}^3a^4f^4)\right)^{1/2}}\right]\frac{df}{dt} \qquad (S10)$$

where $c_1$, $c_2$, and $c_3$ are constants equal to $8.74\times10^3$, $2.39\times10^4$, and $8.16\times10^4$, respectively.

To get *dn/dt (mol/s)* we can use *df/dt*, the rate of the frequency decay from the linear fit of the membrane frequency versus time data. We then normalize the leak rate by dividing the calculated *dn/dt* by the pressure driving force for each of the gases measured to get the leak rate (normalized *dn/dt*) into the graphene-sealed microcavity.

**Additional Membranes Measured**

Three additional membranes where measured, two monolayer and two bilayer samples. The monolayer sample in Fig. S5 ("Mono- 3.4 Å") shows similar behaviour as seen in "Bi- 3.4 Å" of the main text. This monolayer sample was filled with 150 kPa above ambient pressure with pure $H_2$. The pore was not stable and additional measurements could not be taken. The second monolayer sample shown in Fig. S6 was measured using the mechanical resonance scheme presented in the main text. This membrane showed a similar pore instability as the previous sample. The order of the leak rate measurements taken on this membrane were $N_2$ (black), $H_2$ (red), $CO_2$ (green), and $CH_4$ (blue). Next, $N_2$ was measured a second time (cyan) showing a drastic increase in the $N_2$ leak rate. After the repeat of the $N_2$ data, we then introduced $SF_6$, and the results show that the membrane is slowly allowing $SF_6$ to permeate indicating that this pore is larger but similar in size to $SF_6$ (4.9Å)[7]. We attribute this increase in $N_2$ leak rate to etching of the pore during the resonance measurement.

Two additional bilayer membranes from the same graphene flake found in Fig. 1 (containing membrane "Bi- 3.4 Å") of the main text are shown in Fig. S7. Figure S7a is a membrane that has larger pores than that of the sample presented in the main text. The membrane in Fig S7a was damaged before $CH_4$ leak rate data could be taken. Fig S7b is the sample presented in the main text, and Fig S7c shows the leak rate of a membrane that showed molecular sieving of $H_2$ versus $CO_2$ and larger molecules (Ar, $N_2$, and $CH_4$). This suggests that the pore size for the membrane in Fig S7c is between 2.89 Å and 3.3 Å[7].

**Comparison to Modeling Results and Effusion**

Jiang et al. simulated transport for two types of pores, a N-terminated one with a ~3 Å size and an H-terminated one with a ~2.5 Å size[8]. Their nominal $H_2$ permeance of 1 mol $m^{-2}$ $s^{-1}$ $Pa^{-1}$ was based on the N-terminated pore at 600 K with a pass through frequency of $10^{11}$ $s^{-1}$ where a 1 bar pressure drop was estimated from their simulation. When discussing the H-terminated (2.5 Å) pore at room temperature Jiang et al states that for $H_2$ the "passing-through frequency" is $10^{9}$ $s^{-1}$. This "passing-through frequency" is lower than the N-terminated by approximately two orders of magnitude for room temperature operation. Thus, we start with 1 mol $m^{-2}$ $s^{-1}$ $Pa^{-1}$ at 600 K and lower it to $10^{-2}$ mol $m^{-2}$ $s^{-1}$ $Pa^{-1}$ to accommodate the fact that our measurements were at room temperature. Then we multiply by the area that Jiang et al. used, which was 187 Å$^2$ (1.87 x $10^{-18}$ $m^2$), to arrive at ~$10^{-20}$ mol $s^{-1}$ $Pa^{-1}$.

To compare to selectivities predicted by the classical effusion model we plotted the leak rate for $H_2$ and $CO_2$ for membrane "Bi- 3.4 Å" and $H_2$, $CO_2$, $N_2$, and $CH_4$ for membrane "Bi- 4.9 Å" and included this in Figure S8 which is a plot of the normalized leak rate versus the inverse square root of the molecular mass of each gas species. Classical effusion predicts that the flow rate through a pore would scale with the inverse square root of the molecular mass and therefore would monotonically increase with increasing inverse square root of the molecular mass. We can also compare the selectivity of $H_2$ to $CO_2$ for both membranes. For classical effusion the selectivity is the ratio of the square root of the molecular masses which is 4.7 for the case of $H_2$ to $CO_2$. For membranes “Bi- 3.4 Å” and “Bi- 4.9 Å” presented in Fig S8 the $H_2$ to $CO_2$ selectivities are 1.7 and 3 respectively. Tables S1, S2, and S3 show the ideal selectivity for “Bi- 3.4 Å”, “Bi- 4.9 Å”, and “Mono- 5 Å”, respectively. This suggests that we are not in the classical effusion regime. Classical effusion requires the pore size to be smaller than the mean free path of the molecule which is ~60 nm at room temperature and ambient pressures. However, we are in a regime where the pore size is much smaller and on the order of the molecule size, therefore it is necessary to consider the molecular size and chemistry.

## Air leaking back into Microcavity

Figure S9 shows air leaking back into a microcavity after all the $H_2$ had rapidly escaped after etching. This is a bilayer sample that was etched in the same manner as the membranes presented in the main text. After etching the sample was filled with 200 kPa of $H_2$ before being imaged. Hydrogen quickly leaks out leaving a near vacuum in the microcavity under the graphene. After 3000 min the deflections changed from -90 nm to -50 nm. This leak rate is consistent with previously measured leak rates for air leaking into an initially evacuated microcavity of similar geometry[3]. This result further suggests that we are measuring the transport thorough the porous graphene for $H_2$ while $N_2$ is diffusing through the silicon oxide substrate.

**Supplementary References:**

1. Koh, Y.K., Bae, M.-H., Cahill, D.G. & Pop, E. Reliably counting atomic planes of few-layer graphene (n > 4). *ACS Nano* **5**, 269-274 (2011).

2. Reif, F. *Fundamentals of Statistical and Thermal Physics*. 651 (McGraw-Hill Book Company: New York, NY, 1965).

3. Bunch, J.S. *et al.* Impermeable atomic membranes from graphene sheets. *Nano Lett.* **8**, 2458-2462 (2008).

4. Hencky, H. Uber den spannungzustand in kreisrunden platten mit verschwindender biegungssteiflgeit. *Z. fur Mathematik und Physik* **63**, 311-317 (1915).

5. Koenig, S.P., Boddeti, N.G., Dunn, M.L. & Bunch, J.S. Ultrastrong adhesion of graphene membranes. *Nature Nanotech.* **6**, 543-546 (2011).

6. Timoshenko, S., Young, D.H. & Weaver, W. *Vibration Problems in Engineering*. 481-484 (John Wiley and Sons, Inc.: New York, 1974).

7. Breck, D.W. *Zeolites Molecular Sieves: Structure, Chemistry, and Use*. 593-724 (Wiley: New York, NY, 1973).

8. Jiang, D.-en, Cooper, V.R. & Dai, S. Porous graphene as the ultimate membrane for gas separation. *Nano Lett.* **9**, 4019-4024 (2009).

**Supplementary Figures:**

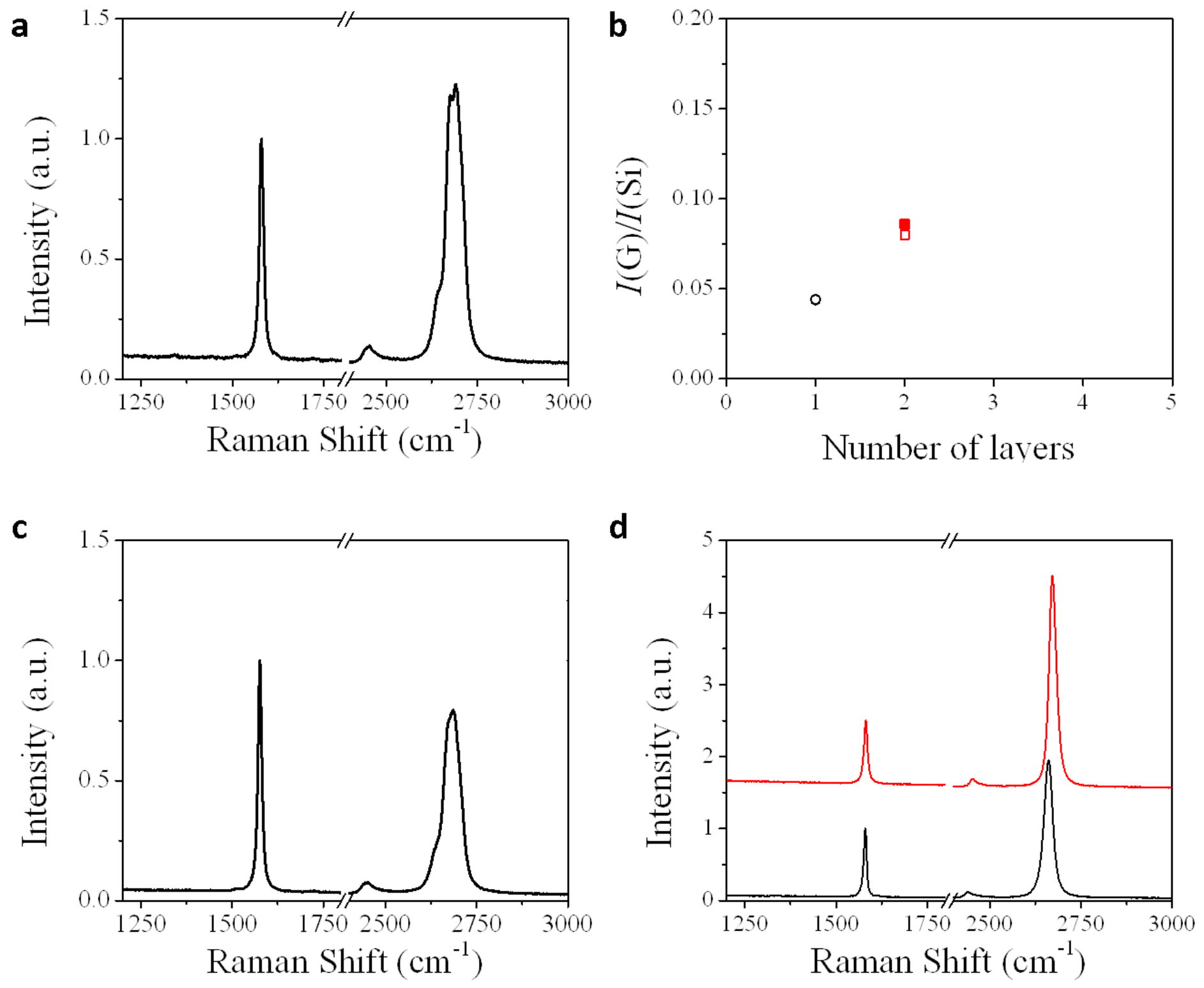

**Figure S1: Raman Spectrum of Graphene Samples**

**(a)** Raman spectrum of graphene flake containing membrane "Bi- 3.4 Å" from the main text taken before etching.

**(b)** *I*(G)/*I*(Si) for flake in (a). The open circle and square were taken from a nearby flake containing both mono and bilayer sections.

**(c)** Raman spectrum of membrane "Bi- 4.9 Å" from the main text before (black) and after (red) etching.

**(d)** Raman spectrum for the monolayer samples presented in the supplementary information. Upper red curve is from the flake containing "Mono- 3.4 Å" from figure S5 of the supplementary information after etching. The lower black curve is for the monolayer membrane presented in Figure S6. Both were taken after etching.

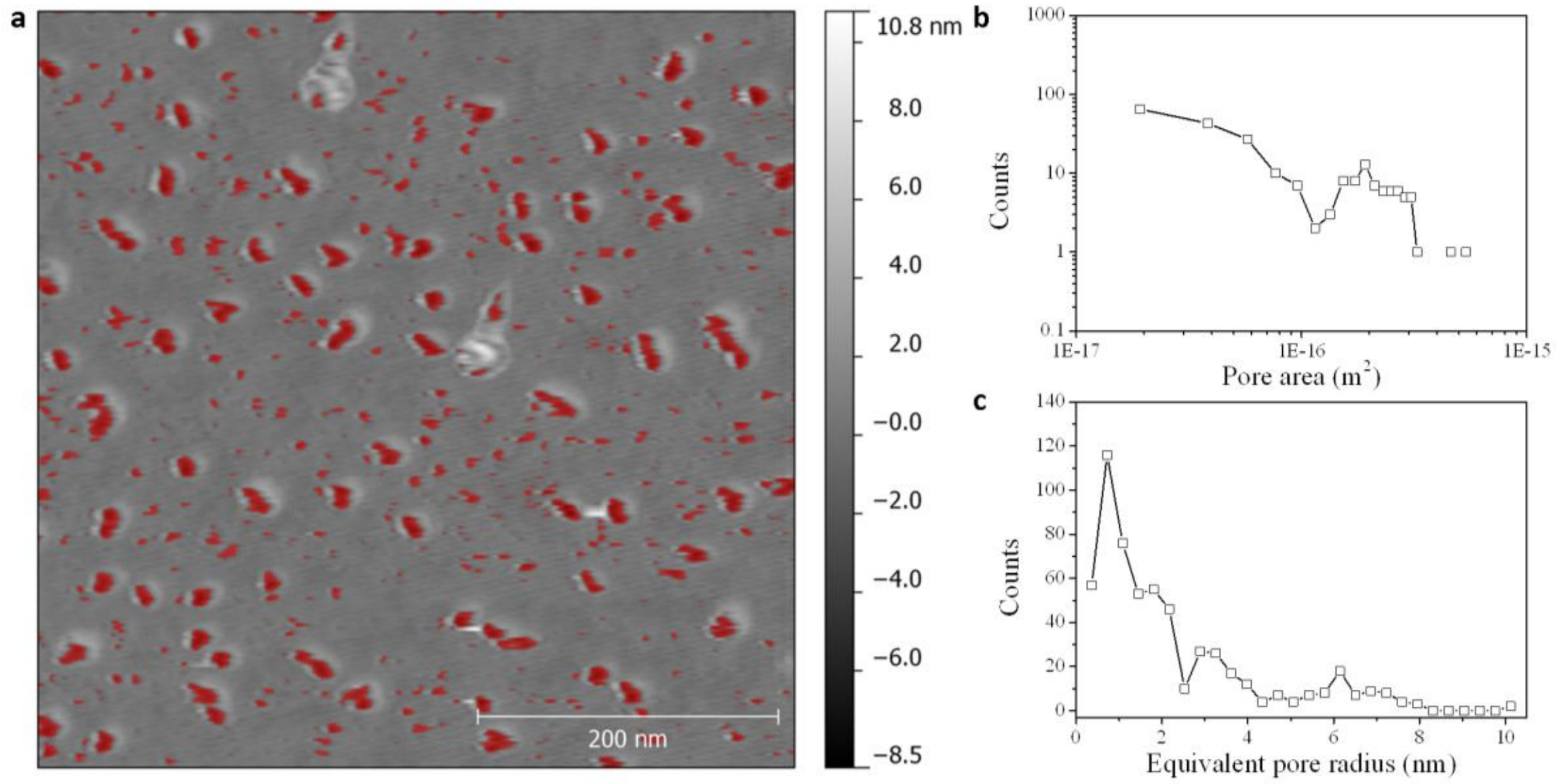

**Figure S2: Visualization of UV etching on suspended graphene**

- **(a)** AFM scan of a membrane etched for a longer time to visualize the pore growth. The red areas are pits created by the UV etching.
- **(b)** Histogram of the number of pores versus the approximate pore area.
- **(c)** Histogram of the number of pores versus the equivalent radius of the pore. **(b)** and **(c)** indicate a nucleation and growth mechanism for pore evolution.

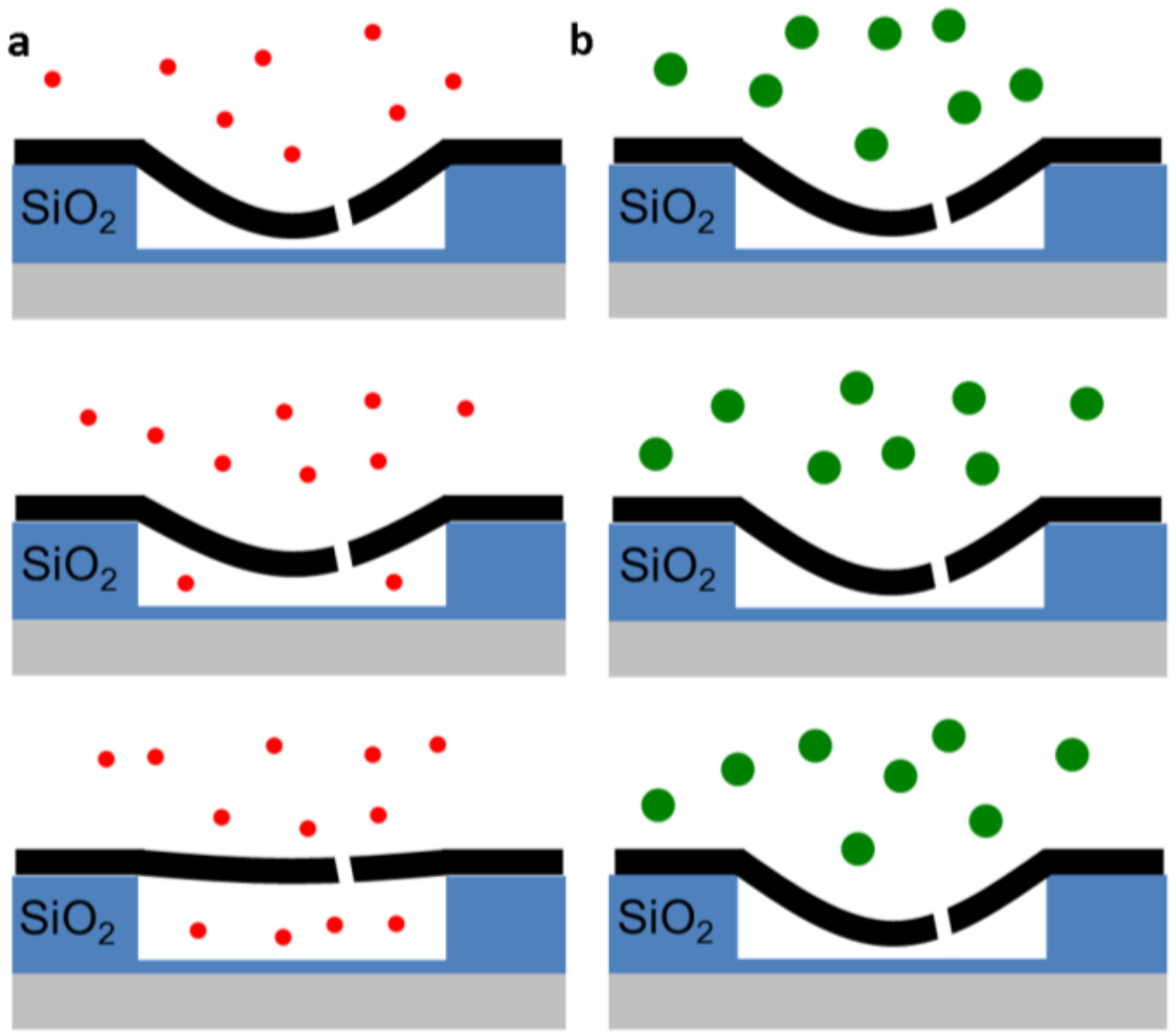

**Figure S3: Schematic of Resonant Frequency Leak Rate Measurements**

**(a)** Schematic of the gas permeation through porous graphene membranes as measured by optical resonance. First the membrane is put in vacuum and the membrane is flat with a frequency of $f_o$ corresponding to zero tension in the membrane. After a pressure of a given gas species is introduced to the vacuum chamber the pressure difference across the membrane will induce tension causing the vibrational frequency to increase. If the gas species kinetic diameter is smaller than that of the pore size (red) it will pass through the pore(s) and the pressure difference will equalize and, therefore, the tension and resonant frequency will decrease with time.

**(b)** If the gas species is larger than the pore size (green), the gas will not pass through the graphene membrane and the tension and resonant frequency will stay constant with time.

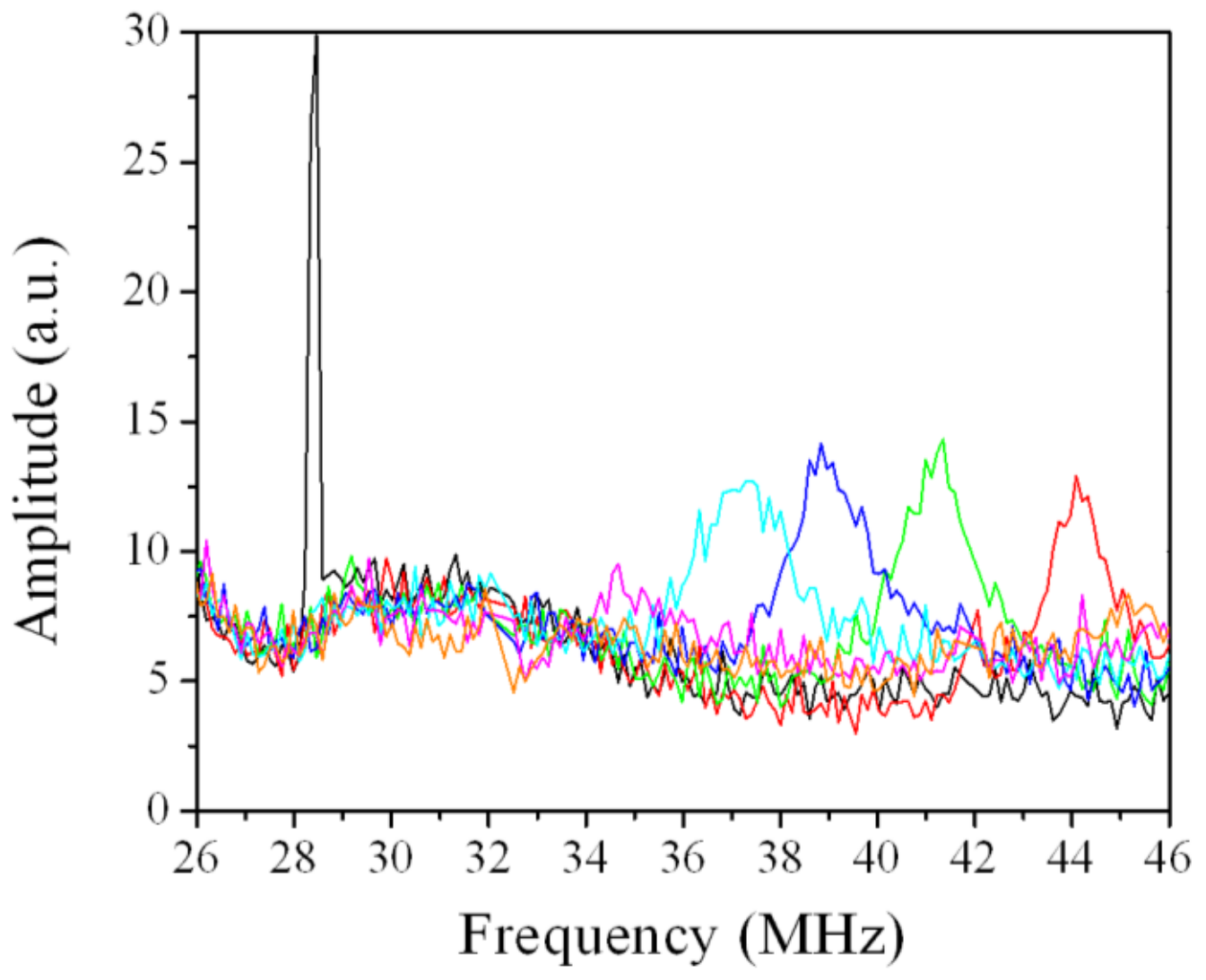

**Figure S4: Sample Resonant frequency curves for $CH_4$**

Amplitude vs drive frequency for 80 torr of $CH_4$. The data corresponds to the frequencies shown in Fig 3 inlay of main text taken at $t$ = 0 s (black), $t$ = 1 s (red), $t$ = 3 s (green), $t$ = 5 s (blue), $t$ = 7 s (cyan), $t$ = 11 s (magenta), and $t$ = 13 s (orange).

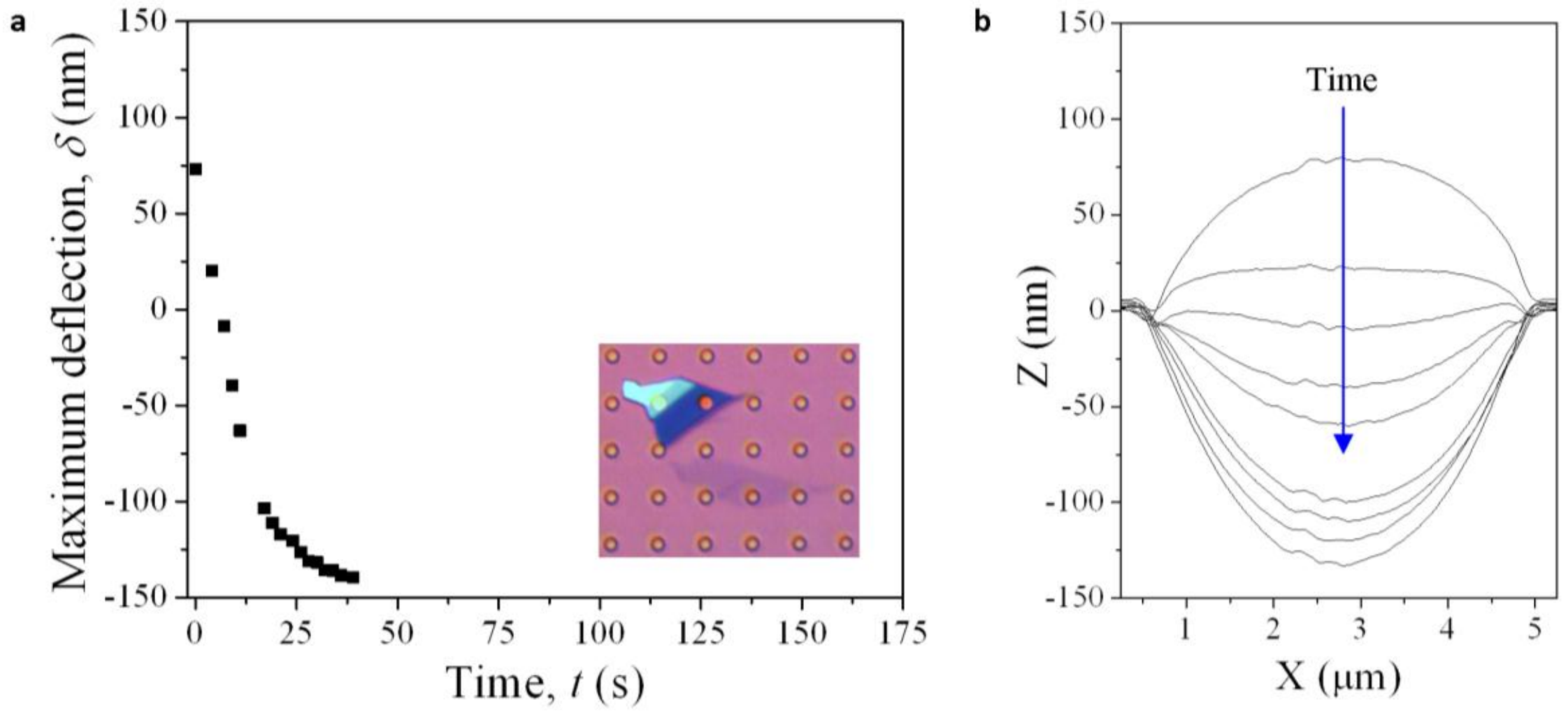

**Figure S5: Monolayer graphene showing selectivity $H_2/N_2$ selectivity**

**(a)** Maximum deflection, $\delta$, vs, $t$ for a monolayer membrane. The rapid decrease in deflection that becomes negative is consistent with the results seen in Fig 1 of the main text. Inlay: optical image of the monolayer graphene membrane covering one well in the substrate.

**(b)** AFM line scans of the membrane in (a) as time passes.

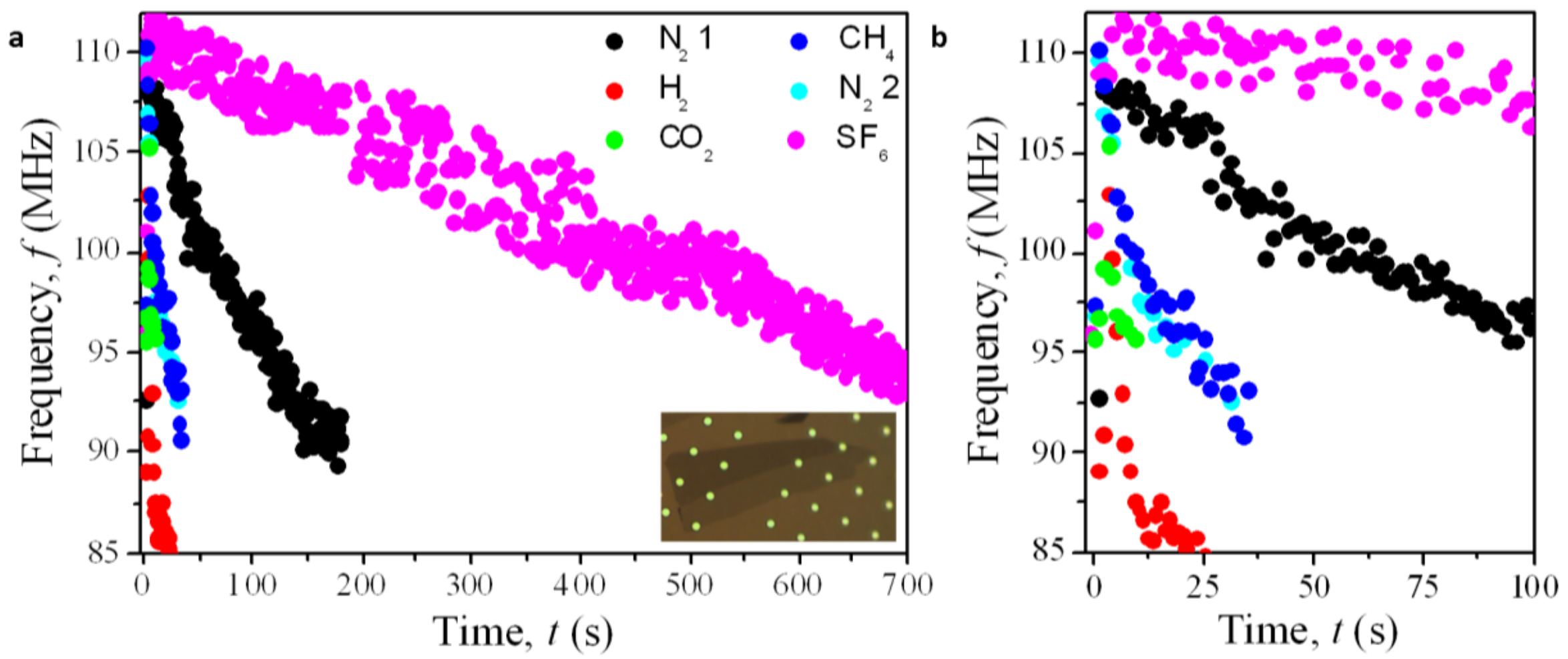

**Figure S6: Monolayer graphene showing $SF_6$ permeation and pore instability**

**(a)** Frequency vs time for $N_2$, $H_2$, $CO_2$, $N_2$, $CH_4$, and $SF_6$, taken in that order.

**(b)** A zoom in of (a). The change in $N_2$ leak rate indicates that the pore(s) in monolayer graphene are not stable and the pore size can change. After the pore was enlarged, the membrane was able to allow $SF_6$ to leak through the membrane.

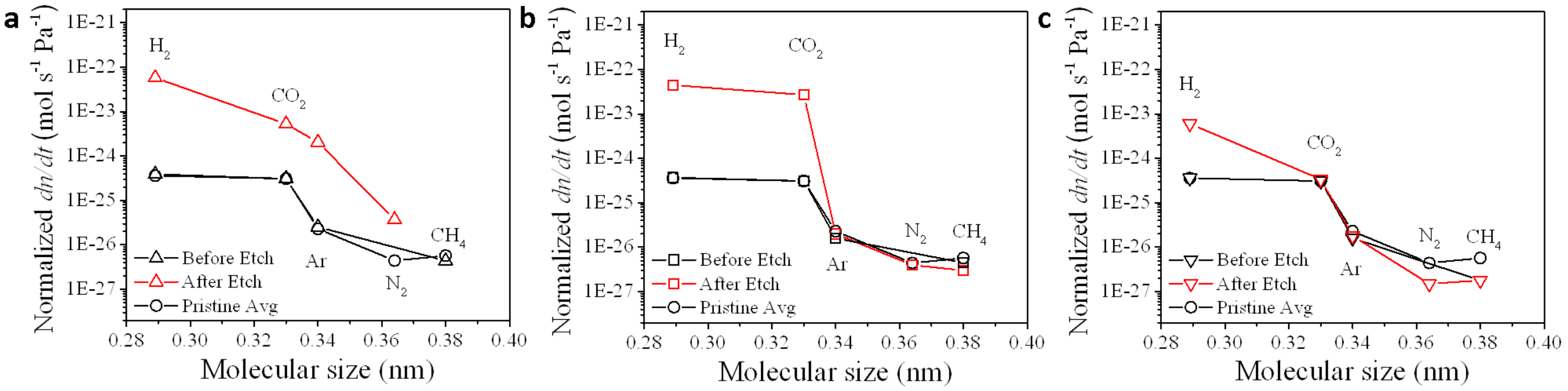

**Figure S7: Additional bilayer membranes measured**

**(a)** Normalized *dn*/*dt* vs. Molecular size showing permeation of all gas species larger than $CH_4$ before and after etching. This membrane was damaged before the $CH_4$ data could be taken.

**(b)** Normalized *dn*/*dt* vs. Molecular size for the membrane "Bi-3.4 Å" before and after etching.

**(c)** Normalized *dn*/*dt* vs. Molecular size for a membrane showing an increase in the leak rate of $H_2$, and no significant increase in the leak rate for $CO_2$, Ar, $N_2$, and $CH_4$. **(a)**, **(b)**, and **(c)** where all from the same graphene flake that can be found in the inlay of Fig 1f.

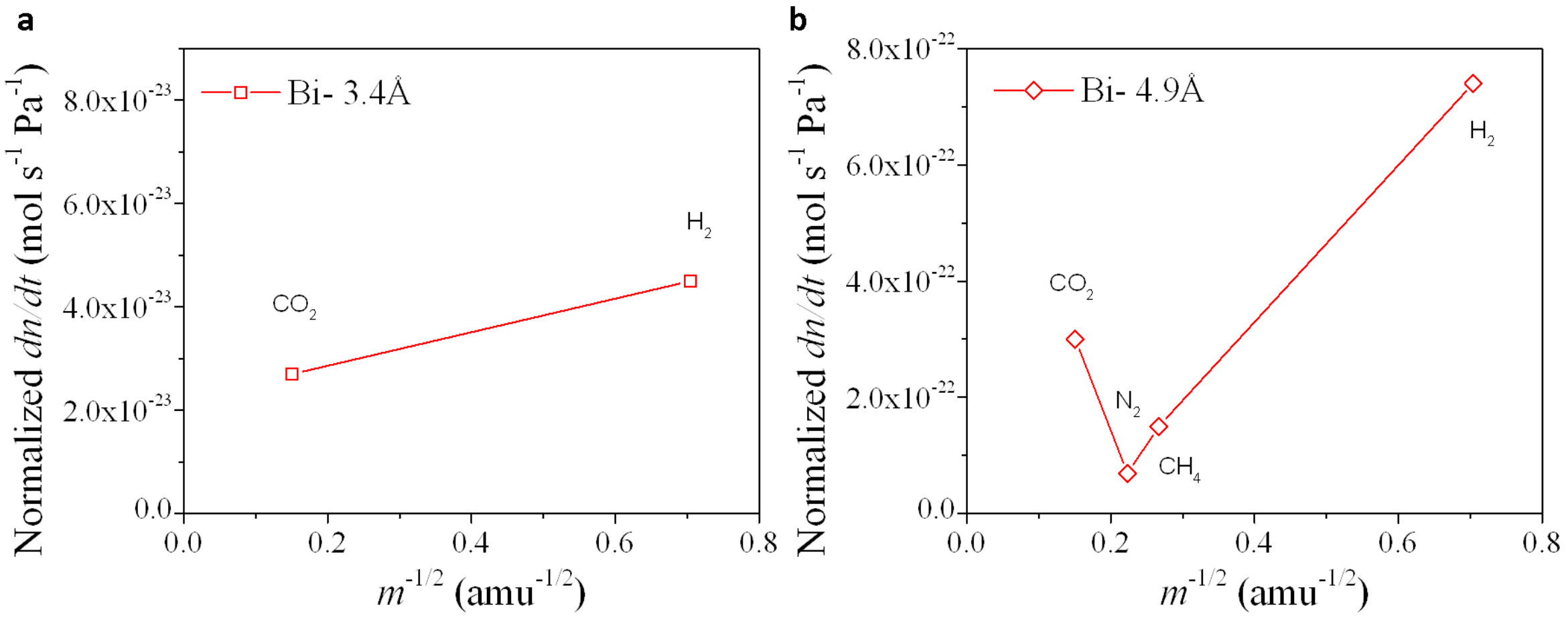

**Figure S8: Comparing Flow Rates to Classical Effusion**

**(a)** Normalized *dn/dt* for membrane "Bi- 3.4 Å" plotted versus the inverse square root of the molecular mass of $H_2$ and $CO_2$.

**(b)** Normalized *dn/dt* for membrane "Bi- 4.9 Å" plotted versus the inverse square root of the molecular mass of $H_2$, $CO_2$, $N_2$ and $CH_4$.

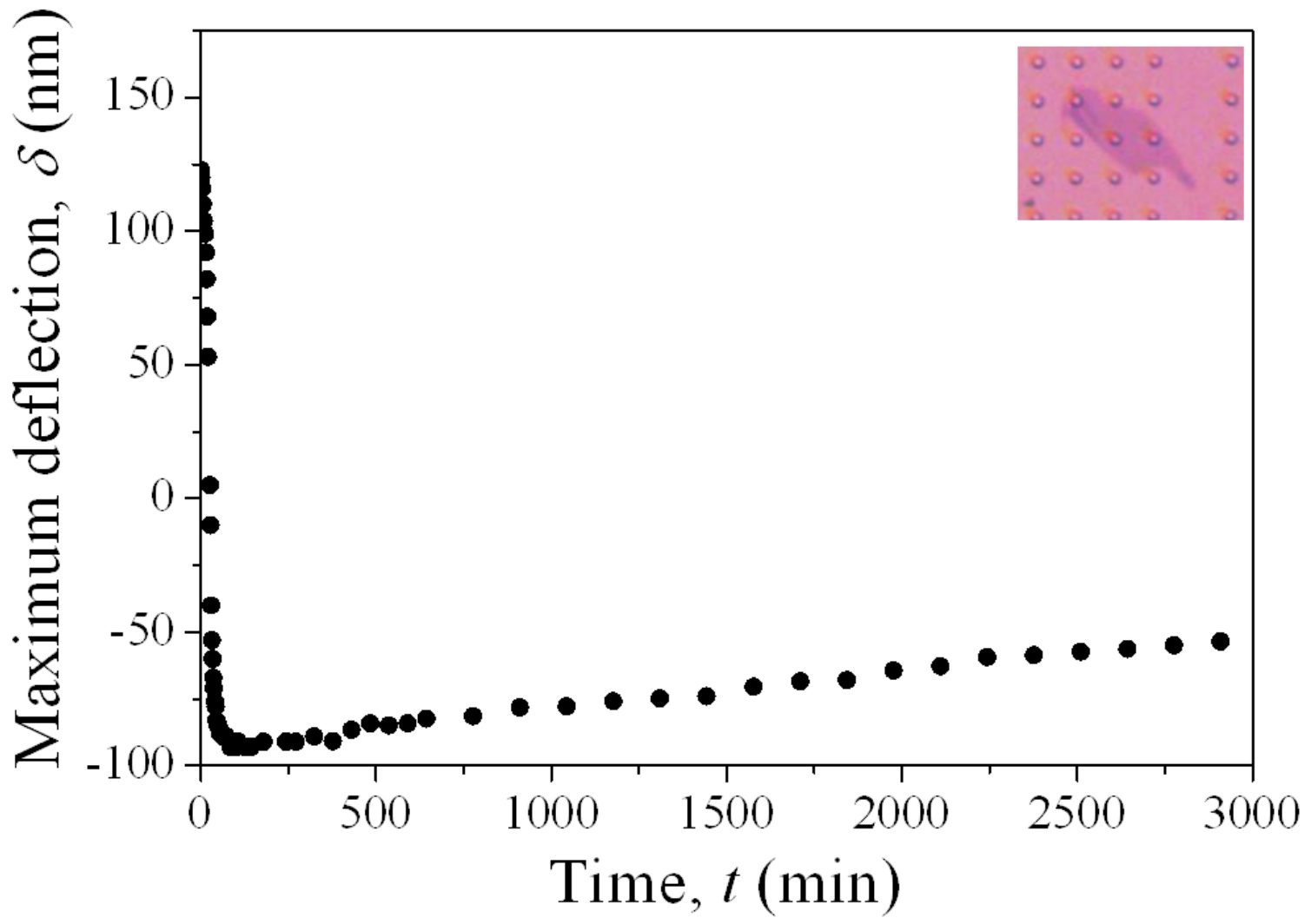

**Figure S9: Air leaking back into Microcavity**

Maximum deflection, $\delta$, vs, $t$ showing the air leaking back into the microcavity after all the $H_2$ has rapidly leaked out through pores created in the graphene. The microcavity was initially filled with 200 kPa of $H_2$. Inlay show the optical image of this sample.

**Supplementary Tables:**

**Table S1: Ideal gas separation factors for membrane "Bi- 3.4 Å"**

| | $H_2$ | $CO_2$ | Ar | $N_2$ | $CH_4$ |
|---|---|---|---|---|---|
| $H_2$ | -- | 1.7 | $2x10^3$ | $10^4$ | $10^4$ |
| $CO_2$ | -- | -- | $5x10^2$ | $7x10^3$ | $9x10^3$ |
| Ar | -- | -- | -- | 5 | 7 |
| $N_2$ | -- | -- | -- | -- | 1.3 |
| $CH_4$ | -- | -- | -- | -- | -- |

**Table S2: Ideal gas separation factors for membrane "Bi- 4.9 Å"**

| | $H_2$ | $CO_2$ | $N_2$ | $CH_4$ | $SF_6$ |
|---|---|---|---|---|---|
| $H_2$ | -- | 3 | 11 | 5 | N/A |
| $CO_2$ | -- | -- | 3.6 | 1.7 | N/A |
| $N_2$ | -- | -- | -- | 0.5 | N/A |
| $CH_4$ | -- | -- | -- | -- | N/A |
| $SF_6$ | -- | -- | -- | -- | -- |

**Table S3. Ideal gas separation factors from membrane "Mono- 5 Å"**

| | $H_2$ | $CO_2$ | $N_2$ | $CH_4$ | $SF_6$ |
|---|---|---|---|---|---|
| $H_2$ | -- | 0.49 | 1.97 | 2.37 | 126 |
| $CO_2$ | -- | -- | 3.95 | 4.89 | 260 |
| $N_2$ | -- | -- | -- | 1.23 | 65.7 |
| $CH_4$ | -- | -- | -- | -- | 53.1 |
| $SF_6$ | -- | -- | -- | -- | -- |